\documentclass{emulateapj}
\slugcomment{{\it Accepted to Nature Physics June 7, 2007. This arxiv.org
version contains more technical details and discussion.}}

\def\xe{x_{\rm e}}
\def\xm{x_{\rm met}}
\def\xmtot{x_{\rm met,tot}}
\def\betat{\beta_{\rm t}}
\def\betarec{\beta_{\rm rec}}
\def\betadiss{\beta_{\rm diss}}
\def\betagr{\beta_{\rm gr}}

\def\keV{{\rm\,keV}}  

\def\s{{\rm\,s}} 
\def\erg{{\rm\,erg}} 
\def\cm{{\rm\,cm}}

\def\mum{\,\mu{\rm m}} 
\def\gm{{\rm\,g}}

\def\AU{{\rm\, AU}}  
\def\K{{\rm\,K}}  
\def\yr{{\rm\,yr}}

\begin{document}

\shortauthors{Chiang \& Murray-Clay}
\shorttitle{Accretion in Transitional Disks}

\title{Inside-Out Evacuation of Transitional Protoplanetary Disks\\
by the Magneto-Rotational Instability}
\author{E.~I.~Chiang\altaffilmark{1,2} and R.~A.~Murray-Clay\altaffilmark{1}}
\altaffiltext{1}{Center for Integrative Planetary Sciences,
Astronomy Department,
University of California at Berkeley,
Berkeley, CA~94720, USA}
\altaffiltext{2}{Alfred P.~Sloan Research Fellow}

\email{echiang@astro.berkeley.edu, rmurray@astro}

\keywords{accretion, accretion disks---X-rays: stars---stars:
pre-main-sequence---solar system: formation---magnetohydrodynamics---planets
and satellites: formation}

\begin{abstract}
  How do T Tauri disks accrete? The magneto-rotational instability (MRI)
  supplies one means, but protoplanetary disk gas
  is typically too poorly ionized to be magnetically active.
  Here we show that the MRI can, in fact, explain observed
  accretion rates for the sub-class of T Tauri disks known as transitional
  systems. Transitional disks are swept clean
  of dust inside rim radii of $\sim$10 AU.
  Stellar coronal X-rays ionize material in the disk rim,
  activating the MRI there. Gas flows from the rim to the
  star, at a rate limited by the depth to which X-rays ionize the rim wall.
The wider the rim, the larger the surface area that
the rim wall exposes to X-rays, and the greater the accretion rate.
  Interior to the rim, the MRI continues to transport gas;
  the MRI is sustained even at the disk midplane by super-keV X-rays that
  Compton scatter down from the disk surface. Accretion is
  therefore steady inside the rim. Blown out by radiation pressure,
  dust largely fails to accrete with gas. Contrary to what is usually assumed,
  ambipolar diffusion, not Ohmic dissipation, limits how much
  gas is MRI-active. We infer values for the transport parameter
  $\alpha$ on the order of a percent
  for the prototypical systems GM Aur, TW Hyd, and DM Tau.
  Because the MRI can only afflict a finite radial column of gas at the rim,
  disk properties inside the rim are insensitive to those outside.
  Thus our picture provides one robust setting for planet-disk
  interaction: a protoplanet interior to the rim will interact
  with gas whose density, temperature, and transport properties
  are definite and decoupled from uncertain initial conditions.
  Our study also supplies half the answer to how
  disks dissipate: the inner disk drains from the inside out by the MRI, 
  while the outer disk photoevaporates by stellar ultraviolet radiation.
\end{abstract}

\section{INTRODUCTION: TRANSITIONAL DISKS AND THE MRI}
\label{sec_intro}

Transitional disks surrounding pre-main-sequence, solar-type stars
are identified by their spectral energy distributions (SEDs).
At 1--10 microns wavelength, their excesses are weaker
than those of classical, flat-spectrum T Tauri disks,
while at longer wavelengths, their fluxes are as strong
as those of any classical T Tauri SED.
The deficit at infrared wavelengths is consistent with
transitional disks having large inner ``holes'' that
are fairly transparent in the continuum.
Figure 1 illustrates the situation.
Outside the hole radius $a_{\rm rim}$, optically thick dust abounds.
Inside $a_{\rm rim}$, only trace amounts of optically thin dust are
present.

Some of the more well-known transitional
systems are listed in Table 1; these
include GM Aur ($a_{\rm rim} \approx 24$ AU;
Calvet et al. 2005, hereafter C05),
TW Hyd ($a_{\rm rim} \approx 1$--4 AU; Calvet et al.~2002, hereafter C02;
Hughes et al.~2007; Ratzka et al.~2007),
DM Tau ($a_{\rm rim} \approx 3$ AU; C05), and
CoKu Tau/4 ($a_{\rm rim} \approx 10$ AU; D'Alessio et al.~2005).
Interferometric measurements verify the existence of
inner disk holes (Hughes et al.~2007; Ratzka et al.~2007; J.~Brown,
personal communication).
Transitional disks may bridge the evolutionary gap between conventional
T Tauri disks that do not have large inner clearings and debris disks that
are entirely optically thin.
Rim radii are too large to represent dust sublimation
fronts, fueling speculation that inside the hole, either
(a) grains have grown too large for the medium to remain optically thick,
(b) planets have consumed or torqued away material, and/or
(c) the inner disk has drained onto the central star by accretion.
This paper concerns possibility (c)---or more accurately,
how accretion in transitional disks is ongoing.

\begin{figure}
\epsscale{1.2}
\plotone{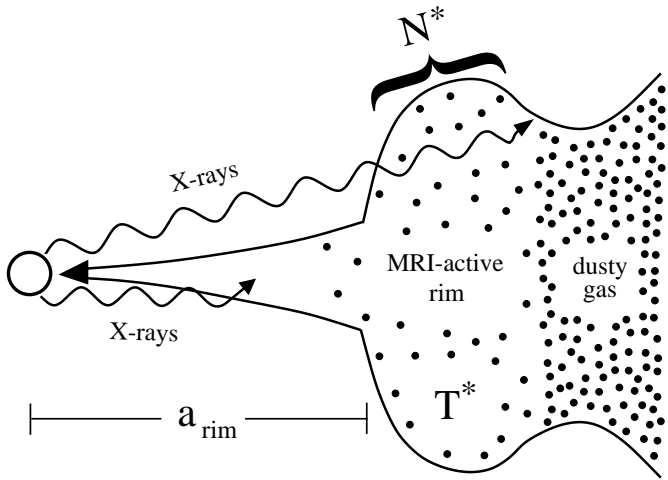}
\caption{Schematic of a transitional protoplanetary disk accreting
by the MRI. Transitional disks
have large rims ($a_{\rm rim} \sim 10$ AU) inside
of which dust is largely absent.
Hard X-ray radiation from the young stellar corona
photoionizes dusty rim gas and heats it to a temperature $T^{\ast}$.
Only a limited gas column $N^{\ast}$ is sufficiently ionized
to be MRI-unstable and drain inwards.  Stellar radiation pressure
clears infalling gas of dust.
}
\label{fig_schem}
\end{figure}

\placefigure{fig_schem}

Transitional disk holes are not empty.
Though nearly devoid of dust, the region interior to $a_{\rm rim}$
often contains accreting gas. Near-ultraviolet excesses, when observed,
imply stellar accretion rates of $\dot{M} \sim 10^{-9} M_{\odot} \,
{\rm yr}^{-1}$ (Muzerolle et al.~2000; C02; Bergin et al.~2004; C05);
see Table 1. We seek here to explain the origin of
such disk accretion.  We propose that gas is leached from the inner
disk rim by
the magneto-rotational instability (MRI), a linear instability that
amplifies magnetic fields in outwardly shearing disks and drives turbulence
(Balbus \& Hawley 1998).
X-rays emitted by the hyperactive corona of the central star irradiate
and ionize the disk rim, activating the MRI there.
Magnetically active gas diffuses inward from the rim to the star.
The inner hole grows as the MRI eats its way out.
The accretion rate interior to the
ever-expanding rim is steady (constant with radius) and entirely set by the
accretion rate established at the rim.

Historically, whether the MRI is relevant to T Tauri disks
has been a prospect fraught with doubt (see, e.g., Hartmann et al. 2006),
as protoplanetary disk gas is
typically too cold and/or too dusty to couple well to magnetic fields.
The following two conditions must be met for the MRI to be viable.
First, magnetic fields must be frozen into whatever
plasma is present. To freeze magnetic flux, the magnetic Reynolds number
must be sufficiently large:
\begin{equation}
{\rm Re} \equiv \frac{c_s h}{D} \approx 1 \left( \frac{a}{{\rm AU}}
\right)^{3/2} \left( \frac{T}{100\K} \right)^{1/2} \left( \frac{\xe}{10^{-13}}
\right) > {\rm Re}^{\ast} \,.
\label{Re}
\end{equation}
Here we follow Fleming et al.~(2000)
in using the gas sound speed $c_s$
and the vertical density scale height $h = c_s/\Omega$
for characteristic
velocity and length scales, respectively; $\Omega$ is
the Kepler orbital frequency; $D$ is the magnetic
diffusivity; $T$ is the gas temperature; $a$ is the disk radius;
and $\xe$ is the fractional electron density.
Numerical simulations by Fleming et al.~(2000)
suggest ${\rm Re}^{\ast} \sim 10^2$--$10^4$,
depending on the geometry of the seed field (but see Pessah,
Chan, \& Psaltis~2007 for a critical
re-analysis of results related to the zero-net flux case;
it is possible only the net vertical and net toroidal
cases have their values of ${\rm Re}^{\ast} \sim 200$--$3000$
well measured).

A second criterion---more often than not neglected in the protoplanetary
disk literature---requires that neutral hydrogen molecules, constituting
the bulk of disk matter, be dragged inward by
accreting plasma (Blaes \& Balbus 1994; Mac Low et al.~1995;
Hawley \& Stone 1998;
Kunz \& Balbus 2004). That is, ambipolar diffusion of the magnetized
plasma out of the (overwhelmingly) neutral medium must not 
defeat the MRI. To couple the neutrals to the ions, a given H$_2$
molecule needs to collide with enough ions within $\sim$$1/\Omega$,
the e-folding time of the instability:

\begin{equation}
{\rm Am} \equiv \frac{x_{\rm e} n \beta_{\rm in}}{\Omega} > {\rm Am}^{\ast}
\,,
\label{Am}
\end{equation}

\noindent where $\beta_{\rm in} \approx 1.9 \times 10^{-9} \cm^3
\s^{-1}$ is the collisional
rate coefficient for ions to share their momentum with
neutrals (Draine, Roberge, \& Dalgarno 1983)
 and $n$ is the number density of hydrogen molecules
(so that $\xe n$ is the number density of ions).
Linear growth rates for the MRI drop
dramatically with the degree to which criterion
(\ref{Am}) is not satisfied (Kunz \& Balbus 2004).
The critical number ${\rm Am}^{\ast}$ of neutral-ion collisions
is measured in numerical simulations to be
about $10^2$ (Mac Low et al.~1995; Hawley \& Stone 1998). To our knowledge,
these pioneering simulations have not been followed up
or verified at higher resolution; therefore we regard
this measurement of ${\rm Am}^{\ast}$,
like the measurement of ${\rm Re}^{\ast}$,
with caution.
Criteria (\ref{Re}) and (\ref{Am}) are more easily
met the more disk gas is ionized; criterion (\ref{Re}) depends
on the fractional ionization, whereas (\ref{Am}) depends
on the total ion density.

Transitional disks furnish an excellent first laboratory
in which to investigate whether the MRI plays a role, if any,
in how T Tauri disks accrete.
Transitional disk holes are
less plagued by sub-micron-sized dust
that threatens to scour the region of plasma.
Furthermore, the rim wall provides an especially simple geometry
for analyzing how X-rays ionize disk material: if the wall is
not severely shadowed from the star by material interior to the rim
(see \S\ref{unresolved}), one need only
treat the 1-D radial transfer of radiation perpendicular to the rim wall.

X-ray driven MRI is not a new idea---it was originally introduced by
Glassgold, Najita, \& Igea (1997, hereafter GNI97)---but our study
applies it for the first time to transitional disks and to identify
the correct criterion for where the MRI can operate under these
conditions.  We argue that the entire region interior to the rim,
including the midplane, is MRI-active. This contrasts with previous
studies of X-ray driven (GNI97; Igea \& Glassgold 1999, hereafter
IG99) or cosmic-ray driven (Gammie 1996, hereafter G96; Sano et
al.~2000) MRI in which accretion is confined to the surface layers of
a disk whose midplane properties cannot be calculated from first
principles.  Moreover, such surface accretion is unsteady (G96). The
accretion rates that we derive here are steady and can be directly
compared to observations. Thus, transitional disks not only provide
insight into the late stages of disk evolution, but also offer the
first clean application of ideas---namely, MRI triggered by
non-thermal ionization processes (G96; GNI97)---pioneered for more
complex systems.

In \S\ref{rim}, we
construct a simple model for the rate of mass flow
from the X-ray illuminated rim. We solve
the equations of ionization and energy balance
to decide how much of the rim satisfies criteria
(\ref{Re}) and (\ref{Am}) and accretes.
In addition to accounting for the usual dissociative
recombination of molecular ions with free electrons (e.g.,
GNI97), our treatment of ionization balance includes
recombination onto dust grains (Umebayashi \& Nakano 1980)
and charge transfer to free atomic metals (Oppenheimer \& Dalgarno 1974,
hereafter OD74;
{}Fromang, Terquem, \& Balbus~2002).
We show how criterion (\ref{Am}), neglected in many studies of
protoplanetary disks (e.g., G96; GNI97;
IG99; Sano et al.~2000; Fromang et al.~2002; Matsumura \& Pudritz 2003),
supersedes criterion (\ref{Re}).
We establish a relation between accretion rate $\dot{M}$
and rim radius $a_{\rm rim}$ and compare to observations.

In \S\ref{steady}, we demonstrate how the MRI continues
to transport gas inward at radii $a \ll a_{\rm rim}$.
Though the midplane of the interior disk is optically thick
to incident X-rays, Compton scattering permits enough X-rays to reach
the midplane (IG99) that the MRI operates there as well.
Accretion is therefore steady from the rim to the star.

In \S\ref{rad}, we study in a preliminary way how stellar radiation
pressure blows dust grains out of accreting gas and can
keep the disk inside the rim relatively transparent.
We make a connection to the small but non-zero optical depth
inferred for the inner disk from SED modelling (C02; C05)
to constrain the abundance of micron-sized grains in the rim.

Our work leaves many issues unresolved and calls attention
to outstanding questions associated with the MRI.
These are discussed in \S\ref{conc}, after a summary of
our principal findings.

\section{ACCRETION FROM THE RIM}
\label{rim}

Ionization of rim gas is maintained by stellar X-rays.\footnote{
Galactic cosmic rays at GeV energies yield ionization
rates of at most $\sim$$10^{-17} \s^{-1}$ per H$_2$ (Spitzer 1978, p.~116).
Actual cosmic ray ionization rates in T Tauri disks are likely
much lower because of shielding by magnetized
T Tauri winds. Even the much weaker solar wind modulates
the cosmic ray flux at Earth by as much as
$\sim$10\% with solar cycle (Marsh \& Svensmark 2003).
We ignore cosmic ray ionization; it is easy to check
that their contribution is negligible compared to ionization by stellar
X-rays.
}
As measured by {\it Chandra}, the X-ray spectra of T Tauri stars
in Orion often have separate soft and hard components,
modelled as emission from thermal plasmas having characteristic
energies $kT \approx 1 \keV$
and $kT \approx 3 \keV$, where $k$ is Boltzmann's constant
(Wolk et al.~2005, see their Figure 12). The luminosities of
the two components are comparable, and depending on the star, 
range from $10^{28}$ to $10^{31} \erg \s^{-1}$.
Soft, sub-keV X-rays are thought to be emitted by shock-heated gas
in accretion columns (Kastner et al.~2002;
Stelzer \& Schmitt 2004), while hard, super-keV photons
may arise from magnetic flares in stellar coronae
(e.g., Wolk et al.~2005).

As the penetrating power of X-rays increases rapidly
with photon energy, we are primarily interested in the hard,
presumably coronal
emission. Unfortunately, we are unable to locate
detailed X-ray spectra at photon energies $E_{\rm X} \gtrsim 3 \keV$
for the prototypical transitional systems discussed in this paper
(GM Aur, TW Hyd, DM Tau, and CoKu Tau/4).
What we glean from the literature is summarized
in Table 1. In the absence of detailed spectra, we assume
for our computations below that
each source emits $L_{\rm X} = 10^{29} \erg \s^{-1}$
at a characteristic energy $E_{\rm X} = 3 \keV$.
This assumption is supported circumstantially by
the aforementioned {\it Chandra} observations of
hard X-ray emission from Orion pre-main-sequence stars.
It is, furthermore, consistent with our findings in Table 1, even for
TW Hyd, whose spectrum is known to be
atypically soft (Stelzer \& Schmitt 2004).

\begin{deluxetable*}{lccccccc}
\tablecaption{Properties of Transitional Disk Systems}
\tablewidth{0pt}
\tablehead{
\multicolumn{8}{c}{}\\
Star & $a_{\rm rim}$ & $a_{\rm rim}$ & $\dot{M}$                    & $\dot{M}$
 & $L_{\rm X,obs}$$^a$     & X-ray band & X-ray \\
     &    (AU)      & refs. &  $(10^{-9} M_{\odot} \yr^{-1})$  &   refs.    &
$({\rm erg} \s^{-1})$  & observed (keV) & refs. \\
}
\startdata
GM Aur & 24 & 1 & 2.7--10 & 1,2 & $10^{30}$ & 0.2--2.4 & 3 \\
TW Hyd & 1--4$^b$  & 4,5,6 & 0.3--0.7$^c$ & 4,7 & $1.4 \times 10^{30}$ &
0.45--2.25 & 8 \\
DM Tau & 3  & 1 & 0.8--2 & 1,2 & $2 \times 10^{29}$ & 0.3--10 & 9 \\
CoKu Tau/4 & 10 & 10 & $< 0.1$ & 11 & NA & NA & 12 \\
\enddata
\tablenotetext{a}{In modelling the hard X-ray spectrum incident upon the rim,
we do not simply use the observed X-ray luminosity $L_{\rm X,obs}$, which
is typically
measured for soft ($E_{\rm X} < 3 \keV$) photons having low penetrating power.
Instead, we assume in our calculations of ionization and thermal balance that
each star emits $L_{\rm X} = 10^{29} \erg \s^{-1}$ at a
characteristic photon energy $E_{\rm X} = 3 \keV$.\\}
\tablenotetext{b}{Controversy exists whether $a_{\rm rim} \approx 4\AU$ (C02;
Hughes et al.~2007) or is nearly $\sim$1 AU (Ratzka et al.~2007) for TW Hyd.
For our analysis we adopt the former result, since it is based on methods
similar to those applied for GM Aur, DM Tau, and CoKu Tau/4. The conclusions of
our paper are unaffected by this choice.\\}
\tablenotetext{c}{Taken to span a factor of 2 to reflect U-band variability
(Muzerolle et al. 2000).}
\tablerefs{
1. C05;
2. Bergin et al.~2004;
3. Strom et al.~1990;
4. C02;
5. Hughes et al.~2007;
6. Ratzka et al.~2007;
7. Muzerolle et al.~2000;
8. Stelzer \& Schmitt 2004;
9. Guedel et al.~2007;
10. D'Alessio et al.~2005;
11. Najita et al. 2007;
12. K\"{o}nig et al. 2001
}
\end{deluxetable*}

Stellar X-rays penetrate the
exposed rim wall to a hydrogen column
density $N$ (see Figure 1). A limited column $N^{\ast}$
will satisfy criteria (\ref{Re}) and (\ref{Am}).
The MRI-active rim
contains mass $M_{\rm rim} \approx 4\pi N^{\ast} a_{\rm rim} h \mu$,
where $h = c_{\rm s}/\Omega$ is the vertical density scale height,
$c_{\rm s} = (kT^{\ast}/\mu)^{1/2}$
is the gas sound speed, $\Omega$ is the Kepler
orbital frequency, $T^{\ast}$ is the kinetic temperature of MRI-active
gas at the rim,
and $\mu \approx 3 \times 10^{-24} \gm$ is the gas mean molecular
weight. This mass flows from $a_{\rm rim}$ to $\sim$$a_{\rm rim}/2$
over the diffusion time $t_{\rm diff} \sim a_{\rm rim}^2 / \nu$,
where $\nu = \alpha c_{\rm s} h$ is the turbulent diffusivity:

\begin{equation}
\dot{M} \approx \frac{3M_{\rm rim}}{t_{\rm diff}} \approx \frac{12\pi \alpha
N^{\ast} a_{\rm rim}^2 (kT^{\ast})^{3/2}}{GM_{\ast}\mu^{1/2}}
\label{M_dot}
\end{equation}

\noindent where $G$ is the gravitational constant and the extra
factor of 3 follows from a more accurate derivation (using
$\dot{M} = 3 \pi \Sigma \nu$, where $\Sigma$ is the disk surface density;
Frank, King, \& Raine 1992).
As measured by numerical simulations of the MRI,
the transport parameter $\alpha$ might be
$\sim$0.01--0.3, depending on the strength and geometry of
the background field (Fleming et al.~2000;
Sano et al.~2004; Pessah et al.~2007; see also \S\ref{unresolved}).
Notice how $\dot{M}$ in equation (\ref{M_dot}) does not depend
explicitly on the surface density profile $\Sigma (a)$
of the disk. Given
observations of $a_{\rm rim}$ and $M_{\ast}$, one need only compute
the rim-specific variables $N^{\ast}$ and $T^{\ast}$, which we do below.

\subsection{Ionization Balance}
\label{ionbal}

Determining $N^{\ast}$ requires that we calculate the degree of ionization
as a function of the radial column $N$
penetrated by X-rays at the rim.
Per H$_2$ molecule, the ionization rate is

\begin{equation}
\zeta = \frac{ L_{\rm X}}{4\pi a_{\rm rim}^2 E_{\rm X}} \eta \sigma_{\rm X}
\exp(-N\sigma_{\rm X})
\label{zeta}
\end{equation}

\noindent
where $\sigma_{\rm X} \approx 4 \times 10^{-24} (E_{\rm X}/3 \keV)^{-2.81}
\cm^2$
is the photoionization cross-section per H$_2$ (assuming solar
abundances)
and $\eta \approx 81$ accounts for the number of secondary ionizations
produced per absorbed 3-keV photon (GNI97).

Freshly ionized H$_2^+$ rapidly converts to molecular ions such as HCO$^+$. 
Most of these molecules dissociate in collisions with electrons,
while some transfer their charge to gas-phase atomic metals
such as magnesium (Fromang et al.~2002; OD74). 
The fractional number abundance $x_{{\rm mol}^+}$
of molecular ions relative to neutral hydrogen molecules is given by the
equilibrium relation

\begin{equation}
{\zeta} = x_{{\rm mol}^+}n (x_{\rm e} \beta_{\rm diss} + x_{\rm met} \beta_{\rm
t})
\label{mol}
\end{equation}

\noindent where $\beta_{\rm diss} \approx 3 \times 10^{-7}
(T/230\K)^{-1/2} \cm^3 \s^{-1}$ and $\beta_{\rm t} \approx 10^{-9}
\cm^3 \s^{-1}$ are rate coefficients for dissociation (Glassgold, Lucas,
\& Omont 1986) and charge transfer (OD74),
and $x_{\rm e}$ and $x_{\rm met}$ are the fractional
number densities of electrons and free neutral metals,
respectively.
To relate the number density $n$ of neutral H$_2$ molecules
to $N$, we spread all
$4\pi Na_{\rm rim}h$ molecules comprising the X-ray penetrated rim
into the volume $2\pi a_{\rm rim}^2 h$ interior to the rim
to estimate that $n \approx 2N/a_{\rm rim}$.

Charge transfer to free metals is important for sustaining the MRI because
ionized metals tend to keep their charge, neutralizing slowly
either by radiative
recombination with electrons or collisions with negatively charged
dust grains (OD74; Umebayashi \& Nakano 1980;
Glassgold et al.~1986):

\begin{equation}
x_{{\rm mol}^+}x_{\rm met}\beta_{\rm t} = x_{{\rm met}^+} (x_{\rm e}\beta_{\rm
rec} + \beta_{\rm gr})
\label{met}
\end{equation}

\noindent where $x_{{\rm met}^+}$ is the fractional abundance of metal
ions, $\beta_{\rm rec} \approx 4 \times 10^{-12} (T/230\K)^{-1/2} \cm^3
\s^{-1}$
is the radiative recombination coefficient,
and $\beta_{\rm gr}$ is the recombination coefficient,
measured per H$_2$, for grains.
We assume the total grain surface area
available for recombination is dominated
by grains of radius $s$ containing a fraction $Z_s$ of the total mass
in gas and dust.
It follows that $\beta_{\rm gr} \approx 3 \times 10^{-20}
(\mu{\rm m}/s) (Z_s/10^{-4}) (T/230 \K)^{1/2} \cm^3 \s^{-1}$.
This estimate, which assumes that every grain is negatively
charged, agrees with that of Umebayashi \& Nakano (1980),
who assume $s \approx 0.1 \mum$ and $Z_s \approx 10^{-2}$.
Though unknown, the factor $Z_s$ is likely to be considerably less
than the maximal value permitted by solar abundance gas, $Z_{\odot}
\approx 10^{-2}$, because of grain growth.\footnote{If grain growth
were so extreme that an individual grain shielded itself from
incident X-rays, then we would have to reduce the X-ray
cross-section $\sigma_{\rm X}$ (which is measured per H$_2$).}
We take as standard values $Z_s = 10^{-4}$ and $s = 1 \mum$.
A constraint on $Z_s / s$ from observations is contained in \S\ref{rad}
on radiation blow-out.

In assuming a single grain size, we
neglect effects introduced by smaller-sized grains.
The surface area for ion recombination may be dominated
by very small grains. Furthermore, because grains
efficiently adsorb electrons, they can, if sufficiently numerous,
completely clear the gas of free electrons and catastrophically
reduce the electrical conductivity. Indeed, this is the
case in the Earth's lower atmosphere, where charges
are carried by aerosols. Fortunately, as we describe
below in \S\ref{acrar}, our results enjoy some margin
of safety from these effects.
The fractional electron densities that we derive,
$10^{-7}$--$10^{-8}$ (compared, say, with terrestrial
ionization fractions $\sim$$10^{-16}$ at sea level), are large enough
that the chemistry is arguably driven by electrons
(but see \S\ref{unresolved}).

Charge and number conservation read, respectively,
\begin{equation}
x_{\rm e} = x_{{\rm mol}^+} + x_{{\rm met}^+}
\label{charge}
\end{equation}
\noindent and
\begin{equation}
x_{\rm met,tot} = x_{{\rm met}^+} + x_{\rm met} \,.
\label{number}
\end{equation}
\noindent
Because free metals are depleted onto grains,
$x_{\rm met,tot}$ is realistically at most $\sim$$3 \times 10^{-6}$,
corresponding
to 1 out of every 10 metals in the free atomic phase (OD74; Umebayashi \&
Nakano 1980).
We consider values for $x_{\rm met,tot}$
ranging from zero (complete
depletion) up to $10^{-6}$.

We combine equations (\ref{mol}), (\ref{met}), (\ref{charge}),
and (\ref{number})
to derive, under certain approximations,
a quartic equation for $\xe$ in terms of
$\zeta/n$, $x_{\rm met,tot}$, and various rate coefficients.
Equations (\ref{mol}), (\ref{met}), and (\ref{charge}) yield
\begin{equation}
\betadiss \xe^2 + \betat \xm \xe- \frac{\zeta}{n} - \frac{\zeta}{n} \left(
\frac{\xm \betat}{\xe \betarec + \betagr} \right) = 0 \,.
\label{rewrite}
\end{equation}
\noindent The first term on the left-hand side dominates the second term when
\begin{equation}
\xe \gg \frac{\betat}{\betadiss} \xmtot \approx 3 \times 10^{-9} \left(
\frac{\xmtot}{10^{-6}} \right) \left( \frac{T}{230\K} \right)^{1/2} \,.
\label{approx}
\end{equation}
\noindent This is always the case for MRI-active rim material,
as will be evident below (see Figures 2 and 3).
Therefore we drop the second term in (\ref{rewrite})
to write
\begin{equation}
\betat \xm = \left( \xe^2 \frac{n\betadiss}{\zeta} - 1 \right) \left( \betarec
\xe + \betagr \right) \,.
\label{end_part1}
\end{equation}

Combining equations (\ref{mol}), (\ref{met}), and (\ref{number}), we have
\begin{eqnarray}
& (\betadiss \xe + \betat \xm) \xmtot  & \nonumber \\
 & = \xm \left[ \betadiss \xe + \betat \xm + \frac{\zeta}{n} \left(
\frac{\betat}{\betarec \xe + \betagr} \right) \right] \,,
\end{eqnarray}
\noindent which simplifies, according to the
same approximation embodied in (\ref{approx}), to
\begin{equation}
\betadiss \xmtot \xe = \xm \left[ \betadiss \xe + \frac{\zeta}{n} \left(
\frac{\betat}{\betarec \xe + \betagr} \right) \right] \,.
\label{end_part2}
\end{equation}
\noindent We solve (\ref{end_part2}) for $\xm$ and substitute into
(\ref{end_part1}) to write
\begin{eqnarray}
& \betadiss \betarec \xe^4 + \betadiss \betagr \xe^3  + \frac{\zeta}{n} \left(
\betat - \betarec \right) \xe^2 \nonumber \\
& - \frac{\zeta}{n} \left( \betagr + \betat \xmtot \right) \xe - \left(
\frac{\zeta}{n} \right)^2 \frac{\betat}{\betadiss} = 0\,.
\end{eqnarray}
\noindent It is safe to ignore $\betarec$ in comparison to $\betat$, so that
\begin{eqnarray}
& \xe^4 + \left( \frac{\betagr}{\betarec} \right) \xe^3 + \left(
\frac{\zeta}{n\betadiss} \frac{\betat}{\betarec} \right) \xe^2 \nonumber
\label{quart}
\\
& - \left( \frac{\zeta}{n\betadiss} \frac{\betat}{\betarec} \right) \left(
\frac{\betagr}{\betat} + \xmtot \right) \xe  \\
& - \frac{\betat}{\betarec} \left( \frac{\zeta}{n\betadiss} \right)^2
=  0  \,. \nonumber
\end{eqnarray}
This final equation is a quartic and not a cubic (OD74)
because we allow for the possibility that nearly all gas-phase
atomic metals might be ionized, via equation (\ref{number}).

\subsection{Thermal Balance}
\label{thermal}

Gas is heated by fast photoelectrons at a rate per H$_2$ molecule of
\begin{equation}
\Gamma = \frac{L_{\rm X}}{4\pi a_{\rm rim}^2} f \sigma_{\rm X} \exp(-N
\sigma_{\rm X})
\end{equation}
where $f \approx 0.5$ is the fraction of X-ray energy deposited into
heat (Glassgold, Najita, \& Igea~2004). Gas cools by ro-vibrational
transitions of CO at a rate per H$_2$ molecule of
\begin{eqnarray}
\Lambda = & 3.3 \times 10^{-25} (T/\K)^{1/2} \frac{\exp(-4292
\K/T)}{1-\exp(-3084 \K/T)} \times \nonumber \\
& x_{\rm CO} (n/{\rm cm}^{-3}) \,\,\erg \s^{-1}
\label{cool}
\end{eqnarray}
(Glassgold et al.~2004, their equations C7--C8).
Here $x_{\rm CO} \approx 10^{-4}$ is the fractional abundance of CO, whose
level populations are assumed thermal.
Other cooling channels---dust-gas collisions and Lyman $\alpha$ emission
from hydrogen---are less important and ignored.
We replace $n$ in (\ref{cool})
with our estimate $2N/a_{\rm rim}$ (see \S\ref{ionbal}),
set $\Gamma = \Lambda$ in thermal
balance, and solve for $T$ as a function of $N$.
This result is necessary for solving the quartic (\ref{quart})
for $\xe$ because rate coefficients depend on $T$.

For our standard parameters for GM Aur ($L_{\rm X} = 10^{29} \erg \s^{-1}$,
$E_{\rm X} = 3 \keV$, $a_{\rm rim} = 24\AU$),
we find that $T$ drops from about 340 to
200 K as $N$ increases from $10^{22}$ to $10^{24} \cm^{-2}$.
Results for $T$ are very similar for TW Hyd and DM Tau.
The insensitivity to input parameters is due to the exponential
sensitivity of the cooling rate $\Lambda$ to $T$.


\subsection{The Active Column and Resultant Accretion Rate}
\label{acrar}

Taking the run of $T$ with $N$ from our simple model of thermal
balance in \S\ref{thermal},
we solve the quartic (\ref{quart}) numerically.
The solution $x_{\rm e}$ is displayed against $N$
in Figure 2 for parameters appropriate to GM Aur,
for various choices of $x_{\rm met,tot}$.
Evidently the presence of free atomic metals raises
the ionization fraction by up to two orders of magnitude.
Results for TW Hyd and DM Tau are similar to within factors of two.

\begin{figure}
\epsscale{1.5}
\plotone{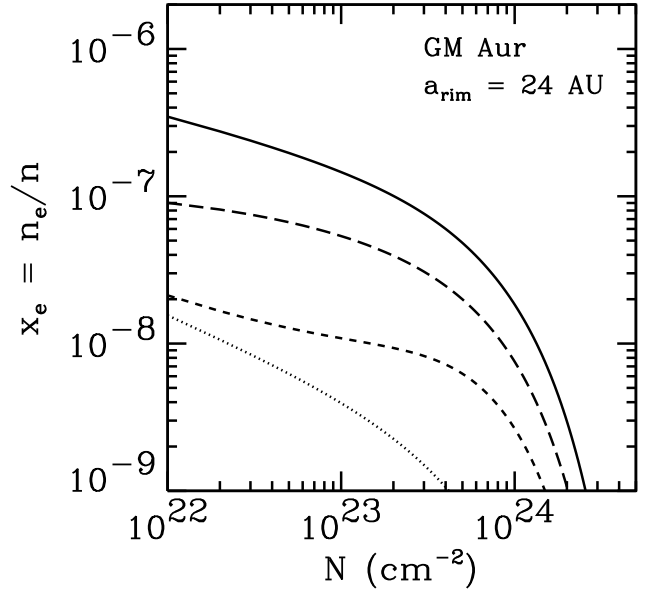}
\caption{
Ionization fraction $x_{\rm e}$ against
column density $N$ penetrated by X-rays at the rim, evaluated for GM Aur
(assuming $L_{\rm X} = 10^{29} \erg \s^{-1}$ and $E_{\rm X} = 3\keV$).
Curves are computed using the quartic equation (\ref{quart})
for $x_{\rm e}$. Results for TW Hyd and DM Tau are similar.
Solid, long-dashed, dashed, and dotted curves correspond
to various choices for the gas-phase atomic metal
abundance $x_{\rm met,tot}$:
$10^{-6}$, $10^{-7}$, $10^{-8}$, and $0$, respectively.
An abundance $\xmtot = 10^{-6}$ corresponds to 1 out of
30 metals in the atomic gas phase.
Free atomic ions, which recombine slowly,
can increase the ionization fraction by up to
two orders of magnitude.
}
\label{fig_xe}
\end{figure}

\placefigure{fig_xe}

{}From $\xe (N)$ we compute ${\rm Am}(N)$,
as shown in Figure 3 for the case of GM Aur.
If ${\rm Am} > {\rm Am}^{\ast} = 100$ strictly determined
the viability of the MRI, then
we would equate $N^{\ast}$ with the maximum value of $N$
for which ${\rm Am} = 100$. However, given uncertainties
in the value of ${\rm Am}^{\ast}$ and in the sharpness of criterion (\ref{Am})
in predicting the non-linear, saturated outcome of the MRI,
only an order-of-magnitude estimate for
$N^{\ast} \sim 5 \times 10^{23} \cm^{-2}$ (the value for
which $\rm Am$ peaks) seems justifiable.
We adopt this same value of $N^{\ast}$ for TW Hyd and DM Tau, for which
very similar results for ${\rm Am}(N)$ obtain.
For $N = N^{\ast} = 5 \times 10^{23} \cm^{-2}$, we find from our thermal
balance model that $T = T^{\ast} \approx 230 \K$.

\begin{figure}
\epsscale{1.4}
\plotone{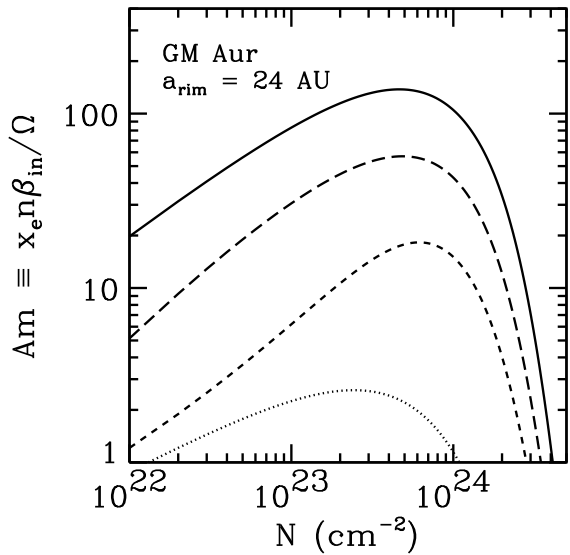}
\caption{
Ambipolar number ${\rm Am}$ against
column density $N$ penetrated by X-rays at the rim, evaluated for GM Aur
(assuming $L_{\rm X} = 10^{29} \erg \s^{-1}$ and $E_{\rm X} = 3\keV$).
Results for TW Hyd and DM Tau are similar.
Solid, long-dashed, dashed, and dotted curves
correspond to various choices for the free atomic
metal abundance $x_{\rm met,tot}$: $10^{-6}$,
$10^{-7}$, $10^{-8}$, and $0$, respectively.
An abundance $\xmtot = 10^{-6}$ corresponds to 1 out of 30
metals in the atomic gas phase (i.e., 29 out of 30 depleted onto grains).
The ambipolar numbers required for accreting
plasma to entrain neutral hydrogen are measured
in numerical simulations to be ${\rm Am}^{\ast} \sim 10^2$.
Roughly speaking, ${\rm Am} \sim {\rm Am}^{\ast}$ for
$N = N^{\ast} \sim 5 \times 10^{23} \cm^{-2}$, 
provided $\xmtot \gtrsim 10^{-7}$.
}
\label{fig_am}
\end{figure}

\placefigure{fig_am}

The ionization fractions,
$x_{\rm e} \sim 10^{-8}$--$10^{-7}$, corresponding to $N \sim N^{\ast}$
are so large that magnetic Reynolds numbers, ${\rm Re}
\sim 10^7$--$10^8$, far exceed the critical values, ${\rm Re}^{\ast}
\sim 10^2$--$10^4$, seemingly required for magnetic flux freezing
(Fleming et al.~2000). Therefore accretion by the MRI in transitional disks
is limited by ambipolar diffusion, not by Ohmic
dissipation.

Electron densities exceed grain densities by 8--9 orders of magnitude,
for our standard $Z_s = 10^{-4}$ and $s = 1 \mum$.
Our neglect of a population of aerosols which can
scour the gas of electrons is arguably safe. The differential grain size
distribution, if it scaled as $s^{-3.5}$, would have to extend
from $s=1 \mum$ all the way down to several angstroms
for the total grain density to be comparable to the electron density.
Our findings
are also insensitive to the uncertain grain recombination coefficient
$\beta_{\rm gr}$,
because metallic ions recombine more readily with electrons than with
grains. For example, increasing $\beta_{\rm gr}$
by a factor of $30$ above our nominal value (such an increase would correspond
to the same grain size distribution mentioned above)
decreases the peak value of ${\rm Am}$ in Figure 3 by a factor of 2,
from 140 to 70.

Armed with $N^{\ast}$ and $T^{\ast}$,
we use equation (\ref{M_dot})
to plot $\dot{M}$ against $a_{\rm rim}$ in Figure 4, adjusting
$\alpha$ as necessary to reproduce observed accretion rates
for GM Aur, TW Hyd, and DM Tau. The offsets
between the different lines drawn for different systems
arise mostly from differences in $M_{\ast}$ and $\alpha$,
since $N^{\ast} = 5 \times 10^{23} \cm^{-2}$ is held fixed
and $T^{\ast}(N^{\ast},a_{\rm rim})$ varies little.
The slopes of the lines
($\dot{M} \propto a_{\rm rim}^2$, approximately)
reflect essentially the increase in the rim wall's surface area
with $a_{\rm rim}$.

\begin{figure}
\epsscale{1.4}
\plotone{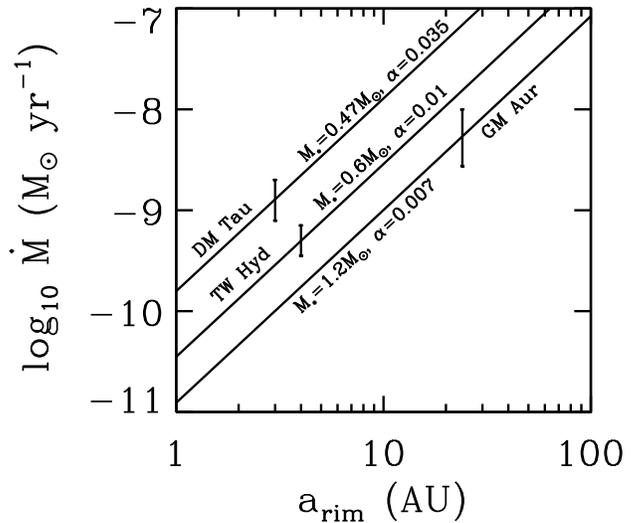}
\caption{
Accretion rates $\dot{M}$ versus rim radii $a_{\rm rim}$.
Points with error bars represent observed data for
GM Aur, TW Hyd, and DM Tau (see Table 1 for references). 
Solid lines are calculated according to equation (\ref{M_dot}),
for $N^{\ast} = 5 \times 10^{23} \cm^{-2}$, $M_{\ast}$ as given by
the literature for each star,
and $T^{\ast}$ as computed as a function of $a_{\rm rim}$ and $M_{\ast}$ for
fixed $N^{\ast}$. Offsets between lines mostly reflect differences
in $M_{\ast}$ and $\alpha$. The accretion rate $\dot{M}$ increases
with $a_{\rm rim}$; the larger the hole, the more surface area
the rim wall exposes to X-rays.
The transport parameter $\alpha$ labelling each curve is chosen
to reproduce the observations.
Fitted values of $\alpha$ range from 0.007 to 0.035,
close to values reported in simulations of the MRI.
}
\label{fig_mdot}
\end{figure}

\placefigure{fig_mdot}

Best-fit $\alpha$'s are 0.007--0.035.
Such values are of order those seen in numerical simulations of the MRI,
depending on the strength and geometry of the background field
(Fleming et al.~2000; Sano et al.~2004; Pessah et al.~2007; see also
\S\ref{unresolved}).\footnote{
Intriguingly, our best-fit $\alpha$'s
resemble those required to fit the light curves of dwarf novae
during their ``cold'' states
($\alpha \approx 0.02$--0.04; Lasota 2001). This resemblance seems sensible
insofar as white dwarf accretion disks
during their quiescent phases and protoplanetary disks
both contain relatively cold, poorly ionized gas.
However, Gammie \& Menou (1998) argue that dwarf nova disks in
quiescence do not support the MRI. These authors consider
the relevant magnetic Reynolds numbers, ${\rm Re} \sim 3700$, too low.
Our own view is that such Reynolds numbers do not proscribe a
weakened form of the MRI (see also Fleming et al.~2000).
}

The transitional system CoKu Tau/4
is not detected in X-rays (K\"{o}nig et al.~2001) and has an unmeasurably
small
accretion rate
($\dot{M} \lesssim 10^{-10} M_{\odot} \yr^{-1}$; Najita et al.~2007;
D'Alessio et al.~2005), facts that at face value are consistent with our
theory.
Possibly CoKu Tau/4 has fewer free atomic metals or---and this is
more readily testable---a softer X-ray spectrum than the other
sources.
If, say, ${\rm Am} \lesssim 10$ at its
rim, then we estimate its 3-keV luminosity $L_{\rm X} \lesssim
10^{28} (10^{-8}/x_{\rm met,tot}) \erg \s^{-1}$.





We expect little accreting material outside $a_{\rm rim}$.
The X-ray heated rim has a vertical thickness greater than that of material
immediately outside. Thus, as depicted in Figure 1, some fraction of the disk
beyond the rim will dwell in the rim's X-ray shadow and be magnetically
inert. Whatever distant material lies outside the rim's shadow and might
be accreting from X-ray irradiation is ignored. See also \S\ref{sum}
for a reason to doubt that the outermost regions of disks accrete.

\section{STEADY ACCRETION IN THE INTERIOR DISK}
\label{steady}

Once dislodged from the rim, gas must still travel up to three decades in
radius, from $\sim$10 to $\sim$0.01 AU,
to reach the stellar surface. We now argue
that X-ray-driven MRI continues to provide transport at $a \ll a_{\rm rim}$.
At $\sim$0.01 AU, the MRI should be well sustained by
thermal ionization. It remains to
decide whether criteria (\ref{Re}) and (\ref{Am}) are satisfied
at the midplane at radii of $\sim$0.1 to $\sim$1 AU.
We need to estimate the temperature and surface density of material
there.

Observed spectra demand that inside the rim, the disk contain
so few grains as to be optically thin at mid-infrared wavelengths (C02; C05).
We can understand this as a consequence both of
the limited number of grains contained within the MRI-active mass
$M_{\rm rim}$, as well as stellar radiation pressure,
which blows out a large fraction of grains having sizes $s < 1 \mu {\rm m}$
(see \S\ref{rad}). Having lost its primary source of
continuum opacity, the gas heats not by incident
starlight, but by accretion (Ohmic dissipation).
Cooling proceeds through line emission, at a rate
that is difficult to estimate because it requires solving simultaneously
for the thermal, chemical, and excitation state of the gas.

We do not attempt to construct a full-blown
thermal model of electrically resistive, line-emitting gas.
Instead, we find it adequate to
estimate the temperature of midplane gas at $a \ll a_{\rm rim}$
in terms of its minimum possible value,
obtained by equating the energy flux from accretion with
that emitted by a blackbody:

\begin{equation}
T \approx 50 \left( \frac{\dot{M}}{10^{-9} M_{\odot} \yr^{-1}} \right)^{1/4}
\left( \frac{a}{{\rm AU}} \right)^{-3/4} \widehat{T} \K \,,
\label{min_T}
\end{equation}

\noindent where $\widehat{T} > 1$ parameterizes our uncertainty.
Mass continuity then implies a vertical hydrogen column density of

\begin{eqnarray}
N_{\perp} \approx & 5 \times 10^{24} \left( \frac{\dot{M}}{10^{-9} M_{\odot}
\yr^{-1}} \right)^{3/4} \left( \frac{a}{{\rm AU}} \right)^{-3/4} \times
\nonumber \\
& \left( \frac{\alpha}{0.02} \right)^{-1} \widehat{T}^{-1} \cm^{-2}
\label{nperp}
\end{eqnarray}

\noindent to the midplane. We have normalized $\alpha$ to the average
of values obtained for GM Aur, TW Hyd, and DM Tau (Figure 4).

Encouragingly, equation (\ref{nperp}) yields
surface densities that are entirely consistent with those
inferred 
by Salyk et al.~(2007) for GM Aur and TW Hyd: at $a \sim 0.2$--0.5 AU,
these authors find $N_{\perp} \sim 0.3$--$3 \times 10^{24} \cm^{-2}$
(H$_2$ surface densities of $1$--$10 \gm \cm^{-2}$
assuming $x_{\rm CO} = 2 \times 10^{-4}$), based on
observations of rovibrational emission from CO. These same observations
suggest that $\widehat{T} \approx 3$ (see their Figure 2),
and we normalize our results below to this value.

To estimate ${\rm Re}$ and ${\rm Am}$ at the midplane,
we first obtain the ionization rate $\zeta$
from IG99, who calculate by Monte Carlo methods
how X-rays incident upon disk surfaces propagate vertically and radially
through protoplanetary disks.
Their calculations account for multiple Compton scattering off bound electrons
and for a thermal spectrum of photon energies.
According to their Figure 3,
for (a) a $kT = 3 \keV$ plasma emitting
$L_{\rm X} = 10^{29} \erg \s^{-1}$, 
(b) $N_{\perp}$ as given by equation
(\ref{nperp}), and (c) $\widehat{T} = 3$,
the ionization rates at the midplane equal $\zeta \approx 10^{-16} \s^{-1}$
at $a = 1$ AU and $\zeta \approx 2.5 \times 10^{-16} \s^{-1}$ at 0.1 AU.

We insert these rates into the quartic equation (\ref{quart})
for $x_{\rm e}$, setting $\beta_{\rm gr} = 0$ because dust
is largely absent in the inner disk.
The quartic yields
$x_{\rm e} \approx 2 \times 10^{-9}$ at 1 AU and
$x_{\rm e} \approx 1 \times 10^{-9}$ at 0.1 AU,
for $x_{\rm met,tot} = 10^{-6}$.
These ionization fractions correspond to ${\rm Re} \sim 2 \times 10^4$
at 1 AU and $900$ at 0.1 AU, values
that satisfy criterion (\ref{Re}), albeit marginally.
Finally, taking $n = N_{\perp} / h = N_{\perp} \Omega / c_s$,
we find that
${\rm Am} \approx 90$ at 1 AU and ${\rm Am} \approx 120$ at
0.1 AU.
These values for ${\rm Am}$, which marginally satisfy criterion
(\ref{Am}), are insensitive to variations in $\widehat{T}$,
which produce changes in $n$ that nearly cancel changes in $x_{\rm e}$.
The margins of safety for ${\rm Re}$ and ${\rm Am}$
would increase by accounting for thermal ionization.

We conclude that the MRI plausibly operates everywhere interior to the rim,
even at the disk midplane,
carrying steadily inward all of the mass drawn from the rim.
Note how the MRI-active vertical columns at 0.1--1 AU,
$N_{\perp} \sim 10^{25} \cm^{-2}$, are larger than the MRI-active
radial column at the rim, $N^{\ast} \sim 5 \times 10^{23} \cm^{-2}$.
This follows from criterion (\ref{Am}), which
assigns importance to the total density $x_{\rm e}n$, not the fractional
density $x_{\rm e}$. Total densities increase dramatically
from the rim to the star.

\section{RADIATION BLOW-OUT OF GRAINS}
\label{rad}

Gas drawn from the rim entrains dust grains.  As the surface density
of inspiralling gas increases towards the star, so, too, does the
surface density of grains. Left unchecked, the increasing
concentration of grains would render the inner disk optically thick at
infrared wavelengths, violating observations.  Here we explore
how stellar radiation pressure can purge
grains from gas (see also Eisner, Chiang, \& Hillenbrand~2006).

Sub-micron-sized grains around solar-type stars
feel an outward stellar radiation force
that just exceeds stellar gravity. Then the time
for such grains to travel from $a$ to $\sim$$2a$ is
\begin{equation}
t_{\rm blow} \sim \frac{1}{\Omega} \left( \frac{ 1/\Omega }{t_{\rm stop}}
\right)
\end{equation}
\noindent where
$t_{\rm stop} \sim \rho_{\rm p} s / (c_{\rm s} \mu n) < 1/\Omega$
is the time for grains of radius $s$ and internal density
$\rho_{\rm p} \approx 1 \gm \cm^{-3}$ to attain terminal velocity according
to the Epstein gas drag law, which applies here because the grain size
is smaller than the collisional mean free path in gas
(e.g., Weidenschilling 1977).
Radiation blow-out is faster
than aerodynamic drift (the latter caused by radial pressure gradients
in gas; e.g., Alexander \& Armitage 2007)
by $(a/h)_{\rm rim}^2 \sim 10^2$, and so we ignore the latter.

We compare $t_{\rm blow}$ to $t_{\rm diff} \sim a^2/\nu$, the time
for gas to diffuse from $a$ to $\sim$$a/2$. At the rim of
the disk of GM Aur, the timescales match by coincidence:
$t_{\rm blow,rim} \sim t_{\rm diff,rim} \sim 1 \times 10^5 \yr$.
This indicates that about half of the dust in $M_{\rm rim}$ is expelled,
leaving the other half entrained with accreting gas and
spread between $a_{\rm rim}$ and $\sim$$a_{\rm rim}/2$.
The geometric optical
depth of entrained dust, measured perpendicular to the midplane, is
\begin{equation}
\tau_{\rm rim} \approx \frac{2N^{\ast}\mu Z_s}{\rho_{\rm p}s} \left.
\frac{h}{a} \right|_{\rm rim} \sim 0.5 \left( \frac{Z_s}{10^{-4}} \right)
\left( \frac{\mu{\rm m}}{s} \right) \,.
\label{tau}
\end{equation}

We can try to check our estimate for $\tau_{\rm rim}$ against
observations of the mid-infrared SED.
Because radiation at a wavelength of $10\mum$ originates
from grains having temperatures of $\sim$300 K, observed $10\mum$
spectra of GM Aur constrain the optical depth in grains at $\sim$1 AU but
not near the rim at 24 AU, where starlight-heated grains would
have colder temperatures of $\sim$100 K.\footnote{X-ray heated
gas and starlight-heated grains generally have different temperatures.}
Therefore we cannot directly
compare our calculated $\tau_{\rm rim}$ with
$10\mum$ observations for GM Aur.

For the case of TW Hyd, however, we can more easily make
this comparison, because its rim is located at $\sim$4 AU.
Scaling to the parameters of that system, we find
$t_{\rm blow, rim} \sim 5\times 10^3 \yr$
and $t_{\rm diff,rim} \sim 2\times 10^4 \yr$,
which suggests that somewhat
more than half of the dust in $M_{\rm rim}$ is
expelled. Reducing our estimate in (\ref{tau}) by an additional factor of 2,
and accounting for the smaller aspect ratio $(h/a)_{\rm rim}$,
we estimate $\tau_{\rm rim} \sim 0.1$ for TW Hyd.
Observationally,
the vertical optical depth of dust interior to the rim
of TW Hydra's disk
is $\tau_{10} \approx 0.05$ at a wavelength of $10\mum$ (C02).
This is essentially the same
as the geometric optical depth for micron-sized grains
because of the silicate resonance band at $10\mum$ wavelength.
Thus, our crude estimate of $\tau_{\rm rim} \sim 0.1$, based on
assumed values for $Z_s = 10^{-4}$ and $s = 1 \mum$,
is within a factor of two of the observed optical depth.
Such approximate agreement we consider acceptable,
and helps to justify our choice earlier for $Z_s/s$ in calculating
the grain recombination coefficient $\beta_{\rm gr}$ (\S\ref{ionbal}).

To satisfy the observation that the disk remain
optically thin at $a \ll a_{\rm rim}$,
the order-half of rim dust that
lies between $a_{\rm rim}$ and $a_{\rm rim}/2$
must fail to penetrate inside $a < a_{\rm rim}/2$.
Radiation pressure ensures that dust does not continue to
inspiral, provided
that $t_{\rm blow}/t_{\rm diff} \propto n T^{3/2} a^{5/2}$
decrease with decreasing $a$. This seems likely to obtain
because $T$ should drop sharply just inside $a_{\rm rim}/2$
once midplane gas becomes too dense to be heated effectively by X-rays.
Unfortunately, a conclusive statement cannot be made without
constructing a model that smoothly bridges conditions at $a \sim a_{\rm rim}$
with conditions at $a \ll a_{\rm rim}$. This would require knowing
in detail how gas heats and cools at all radii.

\section{CONCLUDING REMARKS}
\label{conc}

We summarize our results in \S\ref{sum} and
discuss topics for future research in \S\ref{unresolved}.

\subsection{Summary}
\label{sum}

How do T Tauri disks accrete? We have answered
this longstanding question for the sub-class of objects known as transitional
systems, disks having AU-sized or larger inner holes cleared of
dust. Super-keV X-rays, emitted
by the young star as a consequence of its enhanced dynamo,
irradiate the inner disk rim. A portion of the rim
is sufficiently ionized that it becomes unstable to the magneto-rotational
instability (MRI) and diffuses inward. The MRI, sustained
even at the disk midplane by hard X-rays, transports gas
steadily from the rim all the way to the stellar surface.
We infer transport parameters $\alpha$ on the order of a percent;
such values are indeed yielded by the MRI, according
to contemporary numerical simulations.

We feel the following features of our theory deserve emphasis.

\begin{enumerate}
\item {\it Accretion rate vs.~hole size.}
All other factors (e.g., X-ray luminosity, stellar mass,
free atomic metal abundance) being equal, the accretion
rate $\dot{M}$ should increase nearly as the square of the rim
radius $a_{\rm rim}$, simply from the increased surface
area of the rim wall that is exposed to stellar X-rays.
Late-breaking
data (C.~Espaillat, personal communication) for the
transitional disks CS Cha
($\dot{M} \approx 1.2 \times 10^{-8} M_{\odot} \yr^{-1}$,
$a_{\rm rim} \approx 43 \AU$)
and SZ 18
($\dot{M} \approx 1.7 \times 10^{-9} M_{\odot} \yr^{-1}$,
$a_{\rm rim} \approx 14 \AU$) appear to support this trend of $\dot{M}$ with
$a_{\rm rim}$ as plotted in our Figure 4.
Of course, this trend is expected to apply
only for transitional disks for which $a_{\rm rim} \gtrsim 1$ AU (see the last
issue raised in \S\ref{unresolved}).

\item {\it Ambipolar diffusion vs.~Ohmic dissipation.} Often
it is assumed that Ohmic dissipation controls where the MRI
can and cannot operate. This criterion, embodied in equation
(\ref{Re}) through the magnetic
Reynolds number $\rm Re$, assigns importance to $\xe$, the ratio
of the density of the most mobile charge carriers (electrons) to that
of resistive species (neutral H$_2$ molecules). Large values
of ${\rm Re} \propto \xe$ imply that magnetic flux is
frozen into plasma, enabling the background shear flow to amplify
magnetic fields.

We have shown instead that ambipolar
diffusion, not Ohmic dissipation, dictates the boundaries
of MRI-active zones. Magnetic flux can be perfectly frozen
into plasma; yet if the neutrals do not couple mechanically
to the plasma, the neutrals would be immune to the MRI. Protoplanetary
disk gas is overwhelmingly neutral. Thus, as equation (\ref{Am})
states, a given neutral H$_2$ molecule must collide with
as many (momentum-bearing) ions as possible within a dynamical
time for the MRI to afflict protoplanetary disk gas.
Ambipolar diffusion limits the accretion flow from the rim,
and is only marginally defeated at the midplane far inside the rim.

While we have demonstrated that criterion (\ref{Am}) outweighs
(\ref{Re}) only for transitional disks, we 
suspect the same conclusion holds for conventional T Tauri disks as well.
For example, consider Figure 7 of IG99, which seeks
to describe the depth of the MRI-active layer in disks
without holes. At, say, $a = 1 \AU$, at
the bottom of the supposedly active layer of their model,
the number density is
$n \sim N_{\perp} / h \sim 2 \times 10^{13} \cm^{-3}$. By IG99's
equation (23), the ionization
fraction there is $x_{\rm e} \sim 4 \times 10^{-14}$ (for their standard
$\alpha = 1$). It follows that
${\rm Am} = x_{\rm e} n \beta_{\rm in} / \Omega \sim 10^{-2}$.
This falls four orders of magnitude short of the critical
${\rm Am}^{\ast} \sim 10^2$ seemingly demanded by
numerical simulations (Mac Low et al.~1995; Hawley \& Stone 1998).
Therefore we disagree with IG99's determination
of the depth of the active layer; their layer does not satisfy
criterion (\ref{Am}).

Because the ambipolar diffusion criterion (\ref{Am}) assigns importance
to the total ion density rather than the fractional electron density,
it is hardest to satisfy in the most rarefied
regions of disks, either at large vertical heights or large
radial distance. This is just opposite to what has been
concluded in the literature, that MRI-active zones are restricted
to the surfaces and the outermost portions of disks.

\item {\it Robustness.} Disk properties inside the rim
are insensitive to those outside, since
the MRI can only draw a radial column of $N^{\ast} \sim 5 \times
10^{23} \cm^{-2}$ from the rim at any time.
In other words, unlike the usual situation,
estimating $\dot{M}$ does not require
specifying the detailed surface density profile of the disk $\Sigma (a)$;
we need only know those rim-specific variables in equation (\ref{M_dot}):
$a_{\rm rim}$, $\alpha$, $N_{\ast}$, and $T_{\ast}$.
Our picture therefore provides a robust
setting for theories of how planets form and how their orbits evolve
(e.g., Goldreich \& Sari 2003) within transitional disks.
A protoplanet lying interior to the rim will
interact with gas whose density, temperature, and
transport properties are definite and decoupled from
uncertain initial conditions.

\item {\it How disks dissipate.}
Our study also supplies part of the answer to how disks
dissipate. Excepting matter that gets locked into planets, the
inner disk drains from the inside out by the MRI, while material
beyond the rim photoevaporates by stellar ultraviolet radiation
(Alexander, Clarke, \& Pringle 2006ab). Indeed, the models
of Alexander et al.~(2006ab) require the disk inside
a few AUs to drain onto the star before ultraviolet
radiation can photoevaporate
the outer disk efficiently. Their analysis did not specify the means
by which the inner disk evacuates.
Our work identifies the mechanism:
it is the MRI, leaching material from the disk rim and causing
the inner hole to grow.
How long the MRI takes to eat its way out
depends on the unknown surface density $\Sigma$ of the original disk.
The clearing time $t_{\rm clear} \sim \Sigma a_{\rm rim}^2 / \dot{M} \sim 10^6
\yr$
if $\Sigma \sim 10^2 \gm \cm^{-2}$ at $a_{\rm rim} \sim 10 \AU$,
comparable to that of the minimum-mass
solar nebula.

\end{enumerate}

\subsection{Unresolved Issues}
\label{unresolved}

Numerous areas exist for improvement.

\begin{enumerate}

\item {\it Radiation blow-out.} We have established that a fraction on
the order of 50\% of the sub-micron-sized grains in the MRI-active rim
gets blown back into the rim by stellar radiation pressure.
The remaining
fraction, which gets entrained to $\sim$$a_{\rm rim}/2$ by the
accretion flow, ultimately must also be blown back
to ensure that transitional disk
holes stay as transparent as observed. To prove this last point
requires understanding in detail how gas heats and cools at
all radii from $a_{\rm rim}$ to the stellar surface, and checking
that $t_{\rm blow}/t_{\rm diff} \propto n T^{3/2} a^{5/2}$
decreases with decreasing $a$.

\item {\it Self-shadowing.} Throughout this work
we have assumed for simplicity
that the accretion disk interior to the rim does not obstruct
the rim wall from stellar X-rays. Yet there must be some
self-shadowing, which if sufficiently severe will convert the problem
we have solved to one involving surface irradiation and the two-dimensional
radiative transfer of X-rays (IG99). Even if this were the case, however,
midplane gas at the rim may still be MRI-active by virtue of X-rays that
Compton
scatter downward from the disk surface (IG99; \S\ref{steady}).

The rim might be tall enough
that a good fraction of its lowermost vertical scale height escapes
being shadowed. In our picture, rim gas is optically thin to X-rays,
and is therefore well-heated by them (see also Glassgold et al.~2004).
This source of heating is not enjoyed by midplane gas
in the inner disk, which is optically thick to X-rays.
Thus we envision a sharp drop in temperature and a concomitant
drop in disk thickness just inside the rim.
As was the case for radiation blow-out, addressing the problem of
self-shadowing requires a global treatment of thermal balance.

\item {\it The value of $\alpha$.} We have estimated
that $\alpha$'s on the order of a percent are required
to explain the accretion rates of GM Aur, TW Hyd, and DM Tau.
Such values are observed in numerical simulations of the MRI
under certain assumptions about the magnitude and geometry of
the seed magnetic field. Are these assumptions justified?
What sets the seed field parameters?
And how far can one trust the simulations themselves?

Using vertically unstratified simulations performed by Sano et al.~(2004),
Pessah et al.~(2007) provide
a remarkably simple fitting formula for $\alpha$, which states
that its saturation value scales as the background $B_z/\sqrt{P}$, where
$B_z$ is the mean strength of an assumed
vertical magnetic field, $P$ is the
time-averaged gas pressure,
and the simulation box length $L$
is set equal to the pressure scale height $h$.
Values of $\alpha \sim 0.01$ correspond to 
$B_z/\sqrt{P} \sim 0.008$, or equivalently
a plasma $\beta \equiv 8\pi P/B_z^2 \sim 4 \times 10^5$.
Is such a background state realistic? Can one determine analogous requirements
based on simulations that assume
a net toroidal field, and ultimately on simulations that include
vertical gravity?

\item {\it The value of ${\rm Am}^{\ast}$.} Is ${\rm Am}^{\ast}$ truly
and exactly 100? What is the origin of this dimensionless
number? How sharp is criterion (\ref{Am}) in predicting
the non-linear outcome of the MRI? According to our present calculations,
criterion (\ref{Am}) for ${\rm Am}^{\ast} \sim 10^2$
is only marginally satisfied at the rim and
interior to the rim. We need to know the transport
parameter $\alpha$ as a function of $\rm Am$ (see also point
3 above) to calculate more precisely
the rate of gas flow from the rim.

\item {\it The value of $\xmtot$.} How can one measure the abundance
of gas-phase atomic metals in transitional disks?
Our finding that the MRI is viable depends on
having more than 1 out of every $\sim$300 metal atoms
such as magnesium remaining in the atomic gas phase.
These free metals provide the ions that collisionally
couple the neutrals to the MRI.

\item {\it Aerosols.} What is the abundance of
macromolecules/grains having sizes $\sim$$10 \AA$ to $\sim$$0.1\mum$
and is it large enough to interfere with the standard
electron-driven chemistry we have presented?
On the optimistic side, transitional disks
are observed to have much less micron-sized dust
than conventional T Tauri disks.
We have calculated that fractional
electron densities in the rim are large enough to withstand,
to some degree, the effects of electron adsorption and ion
recombination onto grains. On the other hand,
radiation pressure cannot purge gas of small aerosols because
of their short aerodynamic stopping times and low
efficiencies for intercepting radiation.

\item {\it Accretion rate vs.~X-ray luminosity.} Any mature theory
of X-ray-driven MRI should predict $\dot{M}$ as a function
of $L_{\rm X}$. But without knowing $\alpha ({\rm Am})$ (see points 3 and
4 above),
we cannot derive this relation with confidence.
At the very least $\dot{M}$ must increase
with $L_{\rm X}$. Is there observational evidence for such a trend?
In fact, $\dot{M}$ is known to increase with $M_{\ast}$
(e.g., Hartmann et al.~2006).
And $L_{\rm X}$ is observed to increase with $M_{\ast}$
(Guedel et al.~2007; Telleschi et al.~2007).
But frustratingly enough, when $\dot{M}$ is plotted against
$L_{\rm X}$, whatever trend might be present is obscured
by large scatter (Telleschi et al.~2007).

A shorter-term project would be to measure the hard,
$E_{\rm X} \gtrsim 3 \keV$ spectra of individual sources.
The penetration depth of X-rays
scales strongly with photon energy, as $E_{\rm X}^{2.81}$.

\item {\it From conventional T Tauri disks to transitional disks.}
We have studied the equilibrium dynamics of transitional disks but have said
little regarding origin. Presumably transitional disks evolve
from conventional T Tauri disks.
Are transitional systems, as a group, older?
The age of TW Hyd is estimated to be approximately 10 Myr, which 
indeed is old for a T Tauri star.

How does the evolution from conventional T Tauri to transition
system take place? Unfortunately, we cannot answer this question
by merely
extrapolating our theory to earlier times. Our theory is designed to explain
systems with large, well-defined holes, and would break down
as $a_{\rm rim}$ approaches the stellar radius, if only because
rim gas would be ionized thermally rather than by X-rays.
Data for conventional T Tauri disks clearly do not fit
the relations plotted in Figure 4; their rim radii are on the order
of $\sim$0.1 AU but their accretion rates range up to
$\sim$$10^{-7} M_{\odot} \yr^{-1}$. The origin of transitional
disks will likely remain a mystery until we solve the
decades-old puzzle of how conventional T Tauri disks,
objects without large holes, accrete (see, e.g.,
Hartmann et al.~2006). The solution might involve some combination
of the MRI plus gravitational instability (e.g., G96).

\end{enumerate}

\acknowledgements
  We thank Al Glassgold, David Hollenbach, Uma Gorti, Eric Feigelson,
  Martin Pessah, Catherine Espaillat, Joanna Brown, and Colette Salyk
  for informative discussions.
  Steve Balbus and an anonymous referee
  provided extensive and insightful criticisms that greatly improved
  the content of our manuscript. This work was
  motivated by lectures prepared by one of us (EIC) for
  the Hebrew University Winter School for Theoretical
  Physics, and supported by grants from the National Science Foundation
  and the Alfred P.~Sloan Foundation.

\end{document}